\documentclass[sigconf,natbib]{acmart}
\usepackage{soul}
\usepackage{siunitx}
\usepackage[inline]{enumitem}
\usepackage{multirow}
\usepackage{caption}
\usepackage{subcaption}

\usepackage[subtle]{savetrees}
\copyrightyear{2024}
\acmYear{2024}
\setcopyright{rightsretained}
\acmConference[SIGIR '24]{Proceedings of the 47th International ACM SIGIR Conference on Research and Development in Information Retrieval}{July 14--18, 2024}{Washington, DC, USA}
\acmBooktitle{Proceedings of the 47th International ACM SIGIR Conference on Research and Development in Information Retrieval (SIGIR '24), July 14--18, 2024, Washington, DC, USA}
\acmDOI{10.1145/3626772.3657707}
\acmISBN{979-8-4007-0431-4/24/07}

\makeatletter
\gdef\@copyrightpermission{
  \begin{minipage}{0.3\columnwidth}
   \href{https://creativecommons.org/licenses/by/4.0/}{\includegraphics[width=0.90\textwidth]{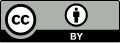}}
  \end{minipage}\hfill
  \begin{minipage}{0.7\columnwidth}
   \href{https://creativecommons.org/licenses/by/4.0/}{This work is licensed under a Creative Commons Attribution International 4.0 License.}
  \end{minipage}
  \vspace{5pt}
}
\makeatother

\makeatletter
\if@ACM@anonymous
  \newcommand{\Bing}{(anonymous)}%
  \newcommand{\anonfootnote}{ (Anonymised for blind review.)}%
\else
  \newcommand{\Bing}{Bing}%
  \newcommand{\anonfootnote}{}%
\fi
\makeatother

\newif\ifdraft
\draftfalse

\newif\ifappendix
\appendixtrue

\begin{document}

\newcommand{\papertitle}{Large~Language~Models~can~Accurately~Predict Searcher Preferences}
\title[\papertitle]{\papertitle}

\author{Paul Thomas}
\orcid{https://orcid.org/0000-0003-2425-3136}
\affiliation{
    \institution{Microsoft}
    \city{Adelaide} \country{Australia}
}
\email{pathom@microsoft.com}

\author{Seth Spielman}
\orcid{https://orcid.org/0000-0002-5089-7632}
\affiliation{
    \institution{Microsoft}
    \city{Boulder} \country{USA}
}
\email{sethspielman@microsoft.com}

\author{Nick Craswell}
\orcid{https://orcid.org/0000-0002-9351-8137}
\affiliation{
    \institution{Microsoft}
    \city{Seattle} \country{USA}
}
\email{nickcr@microsoft.com}

\author{Bhaskar Mitra}
\orcid{https://orcid.org/0000-0002-5270-5550}
\affiliation{
    \institution{Microsoft Research}
    \city{Montreal} \country{Canada}
}
\email{bhaskar.mitra@microsoft.com}

\begin{abstract}

Much of the evaluation and tuning of a search system relies on relevance labels---annotations that say whether a document is useful for a given search and searcher. Ideally these come from real searchers, but it is hard to collect this data at scale, so typical experiments rely on third-party labellers who may or may not produce accurate annotations. Label quality is managed with ongoing auditing, training, and monitoring.

We discuss an alternative approach. We take careful feedback from real searchers and use this to select a large language model (LLM), and prompt, that agrees with this feedback; the LLM can then produce labels at scale. Our experiments show LLMs are as accurate as human labellers and as useful for finding the best systems and hardest queries. LLM performance varies with prompt features, but also varies unpredictably with simple paraphrases. This unpredictability reinforces the need for high-quality ``gold'' labels.

\end{abstract}



\maketitle

\section{Labelling relevance}

Relevance labels---annotations that say whether a result is relevant to a searcher's need---are essential for evaluating information retrieval systems in the ``offline'' or ``Cranfield'' model \cite{sanderson10test}, and as training data for machine-learned systems \cite{liu09learning}.
Labels can come from many sources, but in all cases their quality
can be evaluated by comparing them to some \emph{gold standard} labels~\citep{saracevic2008effects}, from the person who had the need \citep{bailey08relevance}.
Gold labels could originate from a relevance assessor who develops their own query topic, then labels the results. Even better, the originator could be a real searcher who did the query in situ, knows exactly what they were trying to find, and gives careful feedback on what's relevant.

Third-party assessors may disagree with gold because they misunderstand the searcher's preference. If workers are systematically misunderstanding searcher needs (i.e., the labels are biased) this cannot be fixed by getting more data. For example, consider assessors who do not understand which queries are navigational \citep{broder2002taxonomy}. When a first-party searcher wants to navigate to a site, the third-party labels do not reward retrieval of that site. The resulting labels do not help us build a search system that performs well on navigational queries, and this can't be solved by getting more labels from the biased worker pool. Working with human labellers, especially crowd workers, can also lead to other well-documented problems including mistakes, other biases, collusion, and adversarial or ``spammy'' workers \cite{clough13examining,inel23collect,thomas22crowd}. The resulting labels can be low-quality, and using them for training or making decisions will develop a retrieval system that makes similar errors.

\begin{figure}
    \centering
    \includegraphics[width=\columnwidth]{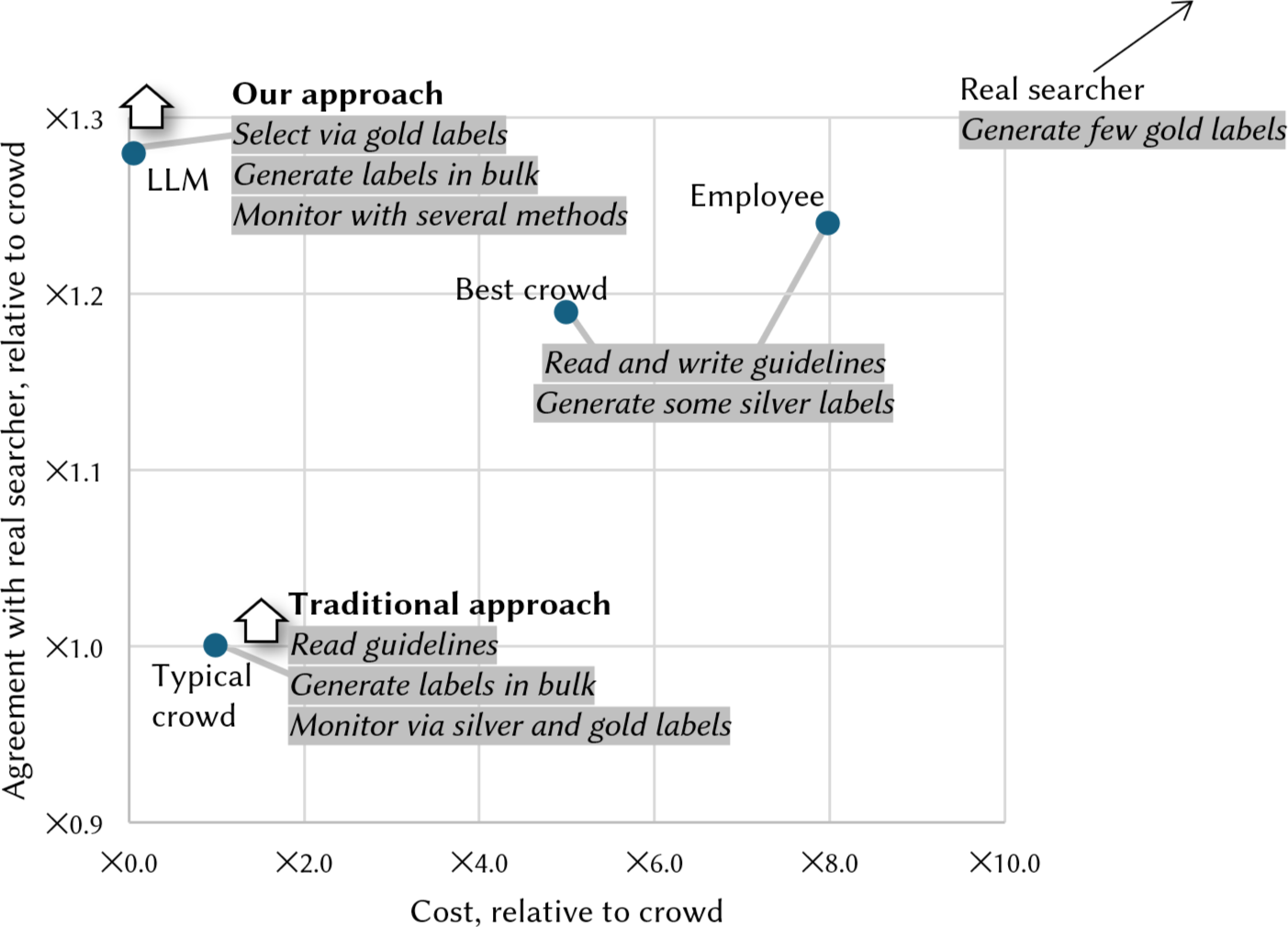}
    \caption{Labelling options discussed in this work, along with the cost and accuracy we see at \Bing.\anonfootnote{}
    A traditional approach uses gold and silver labels to improve crowd workers; we use gold labels to select LLMs and prompts.}
    \label{fig:pareto-space}
\end{figure}

The standard path to obtaining higher-quality labels involves multiple steps (Figure~\ref{fig:pareto-space}). The first is to learn about real searchers through interviews, user studies, direct feedback on their preferences, and implicit feedback on their preferences such as clicks \cite{dumais14understanding}. Studying associated relevance labels, and looking for systematic mistakes, can indicate patterns where labellers are misunderstanding what searchers want. The final step is to train labellers, by reference to guidelines or examples, to minimise future errors: for example, Google uses over 170~pages of guidelines to educate their search quality raters on what makes a good Google result \cite{google22guidelines}. Asking labellers to follow guidelines should lead to improvements in their output, and that improvement can be measured against ground truth that either comes from real searchers (did labellers agree with real searchers?) or against our best understanding of searcher preferences (did labellers agree with examples carefully chosen by experts?).

This paper introduces a new way of reaching very high-quality labels, that match real searcher preferences, by leveraging large language models (LLMs). In practice, LLM performance on any task can vary depending on the choice of model and the wording of the prompt \cite{zhang22tempera,zhou22large}. Our approach is to get a small sample of feedback that perfectly reflects real searcher preferences, because they come from real searchers who did a careful job of giving feedback. We then choose an LLM prompt that generates labels, such that the labels have the best match with first-party ground truth.

For other annotation tasks there is evidence that LLMs can be comparable to crowd workers, using standard metrics such as agreement or correlation \cite{alizadeh23opensource,chiang23llm,gilardi23chatgpt,liu23geval,tornberg23chatgpt}. However, we argue it is more interesting to compare labels to a relatively small set of first-party ground truth, from real searchers. We can then ask how well different labellers do---human or LLM---in generating labels that match real searcher preferences.
Our study shows that \emph{LLM labellers can do better on this task than several populations of human labellers}. Less-trained crowd labellers, who are least knowledgeable about searcher preferences, perform worst as demonstrated by agreement with first-party ground truth. More trained and closely monitored human raters perform better. LLMs, however, perform better on this metric than any population of human labellers that we study. Our results demonstrate the potential for LLMs as a tool for obtaining high-quality relevance labels that match what searchers think.

\section{Labelling relevance with an LLM}

To illustrate these ideas, we have experimented with queries, documents, and labels from TREC-Robust~2004 \cite{voorhees04robust}. Our main question was whether LLMs could replicate the original TREC labels, assigned by expert human assessors.

\subsection{Machinery and data}

TREC-Robust includes 250~topics (each with one canonical query, so ``query'' and ``topic'' are synonymous in what follows)\footnote{One query had no relevant documents. It is included in our analysis but will always score zero, on any metric, using the official labels.}. We took queries from the TREC title field; description and narrative were also included in some prompts, as discussed below.

Official labels were taken from the TREC-Robust qrel file. These labels were assigned by trained assessors, who had also provided the queries and topic descriptions, so although these are not ``real'' in situ search scenarios with a real product, they fit our definition of gold \cite{bailey08relevance}: the person who labelled each document is the single best judge of what the query and topic mean, and what sort of document was responsive. If and when a third-party labeller (human or LLM) deviates from gold, it is considered an error.

The original qrels files had \num{1031} ``highly relevant'' labels, \num{16381} ``relevant'', and \num{293998} ``not relevant''. In the first experiments below we used a stratified random sample of \num{1000} qrels for each label, \num{3000} labelled topic\,:\,document pairs in total. In later experiments, we used all documents returned in Robust 2004 runs at ranks 1--100, where those documents were judged in TREC.

The experiments here used an in-house version of GPT-4 \cite{openai2023gpt4}, representative of the most capable generally-available models at time of writing.
Temperature was set at zero, so the model would select the single most likely output; other parameters were $\mathrm{top}~p=1$, frequency penalty 0.5, presence penalty 0, without stopwords. In early testing and without the prompts below, the model was not able to reproduce TREC documents or qrels with any accuracy above chance.

\subsection{Prompting}

We consider a number of prompt template variants (LLM inputs) which is generally a cheap and fast way to improve quality \cite{karpathy23state}.

\begin{figure}
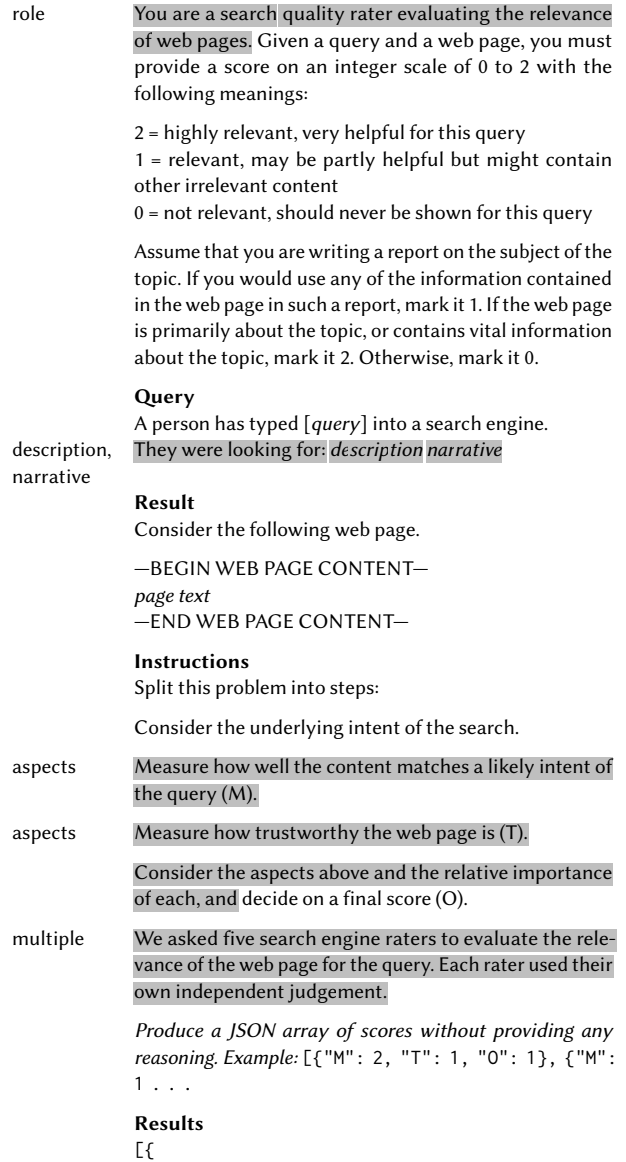

{\sffamily\small {
\sethlcolor{lightgray}
\newcommand{\optionalprompt}[1]{\hl{#1}}
\begin{tabular}{p{0.15\columnwidth}p{0.75\columnwidth}}
role & \optionalprompt{You are a search quality rater evaluating the relevance of web pages.}  Given a query and a web page, you must provide a score on an integer scale of 0 to 2 with the following meanings: \\
\\[-5pt]
& 2 = highly relevant, very helpful for this query \\
& 1 = relevant, may be partly helpful but might contain other irrelevant content \\
& 0 = not relevant, should never be shown for this query \\
\\[-5pt]
& Assume that you are writing a report on the subject of the topic.  If you would use any of the information contained in the web page in such a report, mark it 1.  If the web page is primarily about the topic, or contains vital information about the topic, mark it 2.  Otherwise, mark it 0. \\
\\[-5pt]
& \textbf{Query} \\
& A person has typed [\textit{query}] into a search engine.
\\
description, & \optionalprompt{They were looking for:} \optionalprompt{\textit{description}} \optionalprompt{\textit{narrative}} \\
narrative \\
& \textbf{Result} \\
& Consider the following web page. \\
\\[-5pt]
& ---BEGIN WEB PAGE CONTENT--- \\
& \textit{page text} \\
& ---END WEB PAGE CONTENT--- \\
\\[-5pt]
& \textbf{Instructions} \\
& Split this problem into steps: \\
\\[-5pt]
& Consider the underlying intent of the search. \\
\\[-5pt]
aspects & \optionalprompt{Measure how well the content matches a likely intent of the query (M).} \\
\\[-5pt]
aspects & \optionalprompt{Measure how trustworthy the web page is (T).} \\
\\[-5pt]
& \optionalprompt{Consider the aspects above and the relative importance of each, and} decide on a final score (O). \\
\\[-5pt]
multiple & \optionalprompt{We asked five search engine raters to evaluate the relevance of the web page for the query. Each rater used their own independent judgement.} \\
\\[-5pt]
& \textit{Produce a JSON array of scores without providing any reasoning. Example: \texttt{[\{"M": 2, "T": 1, "O": 1\}, \{"M": 1 \dots}} \\
\\[-5pt]
& \textbf{Results} \\
& \texttt{[\{}
\end{tabular}
}}
\caption{General form of the prompts used in our TREC Robust experiments. \textit{Italicised} words are placeholders, filled with appropriate values. \colorbox{lightgray}{Shaded} text is optional, included in some prompt variants.}
\label{fig:prompt-trecrobust}
\end{figure}

Figure~\ref{fig:prompt-trecrobust} gives an overall schema for the prompts. \textit{Italicised} words are placeholders, which were filled differently for each topic and document, or otherwise varied to match the rest of the prompt. {\sethlcolor{lightgray} \hl{Shaded}} text is optional and was included in some prompt variants.

\paragraph*{Instructions, role} The prompt has four parts.  The first part gives task instructions. These are closely based on instructions given to TREC assessors with two changes: First, the TREC instructions included material on consistency in labels, which is not relevant to an LLM case so was dropped here.  Second, the phrase ``you are a search engine quality rater\dots'' replaces some of the TREC text which discusses the assessors' past experience developing TREC tracks.  Web page quality is a complex notion, but search providers frequently publish hints of what they are looking for. In particular, Google's labelling guidelines use the phrase ``search quality rater'' \cite{google22guidelines}. Half of our prompts therefore include the phrase ``you are a search quality rater evaluating the relevance of web pages'', as a shorthand way to reference both the guidelines (which are generally useful) and surrounding discussion.

\paragraph*{Context, description, narrative} The second part of the prompt gives the query/document pair to be labelled. We always include the query; in some configurations we include a more detailed version from the TREC description and narrative fields; and we give the text of the document itself.

Queries alone are an impoverished representation of an information need, but TREC topics have additional text describing what the query means (description) and which documents should be considered responsive (narrative). For example, for the query \emph{hubble telescope achievements}, the description restates that the query is about achievements of the space telescope since its launch in 1991, and the narrative clarifies that this is about scientific achievement so results that only talk about shortcomings and repairs would not be considered relevant.  In some prompts, we include this text as the ``description'' and ``narrative'' fields.

\paragraph*{Further instructions, aspects, multiple judges} The third part of the prompt restates the task, including the instruction to ``split this problem into steps'' by explicitly considering the searcher's intent as well as the document. This follows observations that ``chain of thought'' or ``step by step'' prompts can produce more reliable results \cite{kojima22zeroshot,wei2023chainofthought} (something we have also observed, informally, in other work).  In some variants, we expanded this to explicitly ask for scores for two aspects---topicality and trust---as well as an overall score.  In some variants, we also ask the model to simulate several human judges and give scores from each.

A straightforward approach, following the TREC guidelines, would be to ask for an overall label for each pair of query\,:\,document.  In past work with human labelling, we have found it more useful to spell out several aspects, and ask for ratings against these, before asking for an overall label.  These extra questions have been useful to help anchor judge assessments, without constraining the final label (i.e. the overall label need not be a simple average of the per-aspect labels).  Similarly, with LLMs there has been demonstrated success with splitting problems into steps with prompts such as ``think step by step'' \citep{kojima22zeroshot}.

Inspired by these ideas, in some prompt variants we explicitly ask for labels over aspects of ``relevance'' as well as for an overall label.  For TREC Robust, we ask for labels for topicality
and for trustworthiness.
There are no further definitions of either aspect.

People naturally vary in their labels, and aggregating several labels for each result can reduce noise and increase sensitivity due to the law of large numbers. In some prompts, we ask the model to simulate several judges, generating the output of five simulated judges from one LLM call. Since the outputs are generated in sequence they are not really independent labellers, but we previously found it useful to generate and aggregate multiple labels in this way, so we include it as a prompt variant here.

\paragraph*{Output} The final part of the prompt specifies an output format, and includes a snippet of JSON to encourage correct syntax.

\vspace{\baselineskip}

\noindent This is a ``zero-shot'' prompt, in that it does not include any examples of the task. \citet{liang22helm} report remarkably mixed results across tasks and models, so it is certainly possible that we could improve with one or more examples; it is also possible we could see some regression. The length of TREC documents means it is hard to include even one entire example, let alone more, and we leave experimentation with one- or few-shot prompts as future work.

Note that we do not claim that this is the best prompt, LLM, nor format; indeed, in Section~\ref{sec:paraphrase} we will see that minor paraphrases can make a material difference. Our interest here is in the range of results we see with a reasonable LLM and prompt (as opposed to the minimal prompts of \citet{faggioli23perspectives} or \citet{liang22helm}), the practical impact of disagreements, and which features of a prompt seem to help or hinder accuracy.

\section{Evaluating the labels}

How are we to choose between labels, or rather between labelling processes?  The main criterion is validity, in particular that labels from any new source should agree with gold labels \cite{faggioli23perspectives}. We can measure this in two ways: by looking at the labels themselves or by looking at preferences between documents. Additionally, labels are typically aggregated to derive query-level or system-level scores, and we may care whether machine labels would lead to similar conclusions at these aggregated levels.

\subsection{Document labels}

The simplest way to evaluate a machine labelling process is to ask: does it produce the same labels as would human labellers?
If the labels are the same for any document, then the machine process can be directly substituted without any quality concerns. 

We can summarise differences between the machine and human labels with a confusion matrix. The labels are on an ordinal scale (not an interval scale), but if we assign scores~0 and~1 to the two levels then we can further compute the mean difference between the human and machine labels. In what follows we report accuracy with the mean absolute error (MAE), where 0~means the two sources always agree on labels and 1~means they are maximally different.

In an earlier study, \citet{faggioli23perspectives} report Cohen's~$\kappa$ between TREC assessors and GPT-3.5 and YouChat LLMs, and we report $\kappa$ here as well. $\kappa$~ranges from 1 (complete agreement) through 0 (agreement only by chance) to $-1$ (complete disagreement). For direct comparison with \citeauthor{faggioli23perspectives} we report $\kappa$ over binarised labels, where partly- and highly-relevant are conflated.

\subsection{Document preference}

Minimising document-level MAE gives us scores which are calibrated across queries, and interpretable for debugging and development. Ranking, however, can use preferences between documents rather than calibrated scores; this is also sufficient for many learning-to-rank algorithms \cite{liu09learning}. Here, it is the relative ordering of any two documents that is important, and we can measure this with pairwise accuracy or AUC: the chance that, given any two documents with a human preference, the model's preference is the same. A score of~1 means the model always agrees with the human's preferences, a score of~0 means they always disagree, and a score of~0.5 is chance alone.

Another consideration is that two scoring schemes may differ in scale and location: for example, one source may give scores 1--5 while another gives 1--10 or 0-99. MAE in this case is misleading, even if there is a completely consistent mapping from one source to another. Pairwise preferences are robust to this sort of difference.

\subsection{Query ordering}

Our primary interest is in generating (and evaluating) labels for documents. However, past work has shown that errors in document labels can be washed out when labels are aggregated to query-level or system-level scores \cite{bailey08relevance}. It is certainly possible that differences in labels are not relevant to query- or system-level evaluations.

In consideration of this we can also order result lists (SERPs) by some metric (e.g. RBP or MAP), according to the labels produced by humans and with regard to some fixed search engine; order the same result lists, on the same metric, according to the labels produced by a model; and ask how similar the two orderings are.

With this query-level analysis we are likely to be looking for queries which do badly (i.e. where a system scores close to zero), so here we measure correlation with rank-biased overlap (RBO) \cite{webber10rbo} after sorting the queries from lowest to highest scores. This means that (dis)agreements about which queries score worst---which queries we want to investigate---count for more than (dis)agreements about those queries that score well.

In our case, since the two rankings are permutations, there is a well-defined lower bound for RBO.
For ease of interpretation we use this minimum to normalise RBO scores into the range 0 to 1, so~0~is an exactly opposite ranking and~1~is an identical ranking. We use set $\phi = 0.9$, corresponding to an experimenter looking (on average) at the first ten queries.

\subsection{System ordering}

The primary use of query:document scores is of course to score a whole system, first by accumulating document scores to query scores then by accumulating query scores to system scores. To see the effect of disagreements between our human and LLM judges, we report RBO over those systems which ran the same queries. Again, since there are a fixed set of systems, we can calculate the minimum RBO score and normalise. An experimenter might look seriously at the top three or four systems, so we set $\phi=0.7$.

\subsection{Ground-truth preferences between results}

An alternative view is that, since human-assigned labels may themselves be biased or noisy, labels should instead accurately predict real searcher preferences.

Evaluating machine labels by their agreement with human labels is useful, because in many situations we can use a large corpus of existing labels. However, it does not speak to the validity of the labels: that is, whether the labels (or a metric derived from the labels) reflects some true searcher experience. If machine labels agree with human labels to (e.g.) 80\%, then the 20\% disagreement might be a fault with the machine, or poor-quality labels from the humans, or some combination. We expand on this idea in Section~\ref{sec:web}.

\subsection{Other criteria}

We can imagine other criteria for choosing a labelling process. These might include cost; time; reliability; scalability; flexibility; and ease of debugging.
In our experience labelling with LLMs is superior to labelling by crowd workers on all these grounds and to labelling by experts
on all except debuggability.

\section{Results}
\label{sec:results}

After running the prompt, we converted the label to a score in $[0, 2]$.  Where we generated multiple labels, the final score is simply the mean.  In keeping with the TREC guidelines, if we prompted for aspects we still considered only the overall label. If the model generated unparseable output, we dropped the result entirely: this happened in~90~out of~\num{96000}~cases.

TREC-Robust included two sets of topics. Topics up to~650 came from earlier editions of TREC, and had only binary relevance judgements (``relevant'' or ``non-relevant''; 1 or 0). Topics 651--700 were developed for the track, and have three-level judgements (adding ``highly relevant'', 2). Our prompts generated a scores from 0~to~2 for all documents, in line with instructions to TREC-Robust assessors for the new topics. Since comparisons are difficult between a three- and a two-level scale, we follow TREC and \citet{faggioli23perspectives} by considering ``relevant'' and ``highly relevant'' together, i.e. by binarising the scores in all cases.

We evaluate the quality of these \emph{labels} (not the documents) in two ways: by comparing the model's labels for each document to the labels from TREC assessors, and by comparing the overall query and system rankings that result. \ifappendix Additional tests are described in the Appendix. \fi

\subsection{Comparing scores}

\begin{table}
  \caption{Performance of the variant prompts of Figure~\ref{fig:prompt-trecrobust}, compared to human labels on a stratified sample of the TREC Robust data. R~=~include role, D~=~include description, N~=~include narrative, A~=~include aspects, M~=~include multiple ``judges''.
  $\star$ marks the best prompt in each case (significantly better than the next-best performer, one-sided $t$ test, $p<0.05$).}
  \small
\begin{tabular}{cccccS[table-format=1.2(2)]S[table-format=1.2(2)]S[table-format=1.2(2)]}
 &&&&& {Document} & {Document} & {Document} \\
 &&&&& {scores} & {scores} & {preference} \\
\multicolumn{5}{c}{Prompt features} & {MAE} & {$\kappa$} & {AUC} \\
\cmidrule(lr){1-5}\cmidrule(lr){6-8}
---&---&---&---&---& 0.34 \pm 0.01 & 0.38 \pm 0.02 & 0.73 \pm 0.01 \\[1em]
 R &---&---&---&---& 0.38 \pm 0.02 & 0.32 \pm 0.02 & 0.71 \pm 0.01 \\
---& D &---&---&---& 0.36 \pm 0.02 & 0.35 \pm 0.03 & 0.72 \pm 0.01 \\
---&---& N &---&---& 0.35 \pm 0.02 & 0.37 \pm 0.03 & 0.73 \pm 0.01 \\
---&---&---& A &---& 0.19 \pm 0.02 & 0.60 \pm 0.03 & 0.82 \pm 0.02 \\
---&---&---&---& M & 0.46 \pm 0.02 & 0.22 \pm 0.02 & 0.65 \pm 0.01 \\[1em]
 R & D &---&---&---& 0.40 \pm 0.02 & 0.30 \pm 0.03 & 0.69 \pm 0.01 \\
 R &---& N &---&---& 0.38 \pm 0.02 & 0.33 \pm 0.02 & 0.71 \pm 0.01 \\
 R &---&---& A &---& 0.21 \pm 0.02 & 0.56 \pm 0.03 & 0.81 \pm 0.02 \\
 R &---&---&---& M & 0.49 \pm 0.02 & 0.20 \pm 0.02 & 0.64 \pm 0.01 \\
---& D & N &---&---& 0.35 \pm 0.02 & 0.37 \pm 0.02 & 0.74 \pm 0.01 \\
---& D &---& A &---& 0.19 \pm 0.01 & 0.59 \pm 0.03 & 0.83 \pm 0.01 \\
---& D &---&---& M & 0.45 \pm 0.01 & 0.24 \pm 0.02 & 0.66 \pm 0.01 \\
---&---& N & A &---& 0.18 \pm 0.01 & 0.62 \pm 0.02 & 0.84 \pm 0.01 \\
---&---& N &---& M & 0.41 \pm 0.02 & 0.29 \pm 0.02 & 0.69 \pm 0.01 \\
---&---&---& A & M & 0.31 \pm 0.02 & 0.42 \pm 0.04 & 0.80 \pm 0.02 \\[1em]
 R & D & N &---&---& 0.37 \pm 0.02 & 0.34 \pm 0.03 & 0.72 \pm 0.02 \\
 R & D &---& A &---& 0.22 \pm 0.01 & 0.53 \pm 0.03 & 0.82 \pm 0.01 \\
 R & D &---&---& M & 0.46 \pm 0.02 & 0.23 \pm 0.02 & 0.66 \pm 0.01 \\
 R &---& N & A &---& 0.20 \pm 0.01 & 0.59 \pm 0.03 & 0.83 \pm 0.01 \\
 R &---& N &---& M & 0.42 \pm 0.02 & 0.28 \pm 0.02 & 0.69 \pm 0.01 \\
 R &---&---& A & M & 0.38 \pm 0.02 & 0.32 \pm 0.02 & 0.78 \pm 0.01 \\
---& D & N & A &---& 0.17 \pm 0.01 & 0.64 \pm 0.02$\star$ & 0.85 \pm 0.01$\star$ \\
---& D & N &---& M & 0.40 \pm 0.02 & 0.31 \pm 0.02 & 0.70 \pm 0.01 \\
---& D &---& A & M & 0.31 \pm 0.01 & 0.42 \pm 0.02 & 0.80 \pm 0.01 \\
---&---& N & A & M & 0.27 \pm 0.02 & 0.49 \pm 0.03 & 0.82 \pm 0.02 \\[1em]
 R & D & N & A &---& 0.19 \pm 0.01 & 0.61 \pm 0.02 & 0.84 \pm 0.01 \\
 R & D & N &---& M & 0.41 \pm 0.01 & 0.29 \pm 0.02 & 0.69 \pm 0.01 \\
 R & D &---& A & M & 0.37 \pm 0.02 & 0.34 \pm 0.02 & 0.80 \pm 0.01 \\
 R &---& N & A & M & 0.33 \pm 0.01 & 0.39 \pm 0.02 & 0.80 \pm 0.01 \\
---& D & N & A & M & 0.26 \pm 0.01 & 0.50 \pm 0.02 & 0.82 \pm 0.01 \\[1em]
 R & D & N & A & M & 0.16 \pm 0.02$\star$ & 0.51 \pm 0.06 & 0.77 \pm 0.03 \\
 \end{tabular}
  \label{tab:results-trecrobust}
\end{table}

Similar to \citet{faggioli23perspectives}, we compare these model-generated scores to scores from the TREC assessors. Table~\ref{tab:results-trecrobust} summarises the models' agreement with human judges, over the \num{3000} query:document pairs, as we manipulate the prompt as above: there is one row for each prompt, identified by which optional features are included. For example, the row labelled ``-{}-N-M'' corresponds to the prompt with \textbf{n}arrative and \textbf{m}ultiple judges, but not \textbf{r}ole statement, \textbf{d}escription, or \textbf{a}spects. For each prompt, we report the three document-level metrics described above, plus a 95\% confidence interval based on 20~bootstraps over documents. The best-performing prompt for each metric is labelled with a $\star$, and these are significantly better than any other ($t$ test, $p<0.05$).

Performance is highly variable as we change the features---that is, the quality of the labelling depends a great deal on the prompt structure or template. For example, Cohen's $\kappa$ varies from as low as 0.20 (prompt ``R-{}-{}-M'') to 0.64 (prompt ``-DNA-''). We need to be accordingly careful interpreting any claim based on a single prompt, especially where that prompt has not been tuned against some existing labels; we also observe this in the variable performance reported in \citet{liang22helm}, for example.

The performance here ($\kappa$ 0.20 to 0.62) compares favourably to that seen by \citet{damessie17gauging}, who re-judged 120~documents from TREC-Robust and saw $\kappa$ of 0.24 to 0.52 for crowd workers, and $\kappa$ of 0.58 for workers in a controlled lab. In particular, 6/32 prompts here to better than 0.58 and only 3/32 do worse than 0.24. Our agreement also compares favourably to  reports from \citet{cormack98efficient}, who labelled TREC ad-hoc documents a second time, using a second group of assessors. From their data we can compute Cohen's $\kappa = 0.52$ between two groups of trained human assessors.

On other data sets, \citet{castillo06spam} report $\kappa = 0.56$ labelling web pages for spam;  \citet{hersh94ohsumed} report $\kappa=0.41$ on relevance in the OHSUMED collection; \citet{agarwal19good} saw $\kappa = 0.44$ for news sentiment; and \citet{scholer13threshold} reported that assessors seeing a document for a second time only agreed with their first label 52\% of the time. \citet{faggioli23perspectives} reported $\kappa$ from 0.26 to 0.40 on binarised labels from TREC-8 and TREC Deep Learning.  \citeauthor{faggioli23perspectives} used another LLM but with relatively simple prompt, reinforcing LLMs' sensitivity to their prompt.

On this metric, at least, we can conclude that with minimal iterations LLMs are already at human quality for this collection and for some prompts. In Section~\ref{sec:web} we will see that, in a web setting, LLMs can perform substantially better than third-party judges.

\subsection{Effect of prompt features}

\begin{table}
  \caption{Performance impact of the optional prompt features in Figure~\ref{fig:prompt-trecrobust}.
  All changes are statistically significant and effects are $\pm 0.005$ at a 95\% CI.}
  \small
\begin{tabular}{rS[table-format=1.2,table-auto-round=true,retain-explicit-plus]}

Feature & {Change in $\kappa$}     \\
\cmidrule(lr){1-2}
Role, R                & -0.041570 \\
Description, D         & +0.013228 \\
Narrative, N           & +0.056793 \\
Aspects, A             & +0.208368 \\
Multiple ``judges'', M & -0.127039 \\
\end{tabular}
  \label{tab:anova-trecrobust}
\end{table}

Table~\ref{tab:results-trecrobust} gives results for 32~prompt templates, made from turning five features on or off. To try to summarise the effect of each feature individually, Table~\ref{tab:anova-trecrobust} reports the effect of each feature on $\kappa$---that is, the effect of including a prompt feature independent of any other features being on or off.

Contrary to our expectations, there is a statistically significant \emph{negative} effect due to role (R) and multiple ``judges'' (M): $\kappa$ decreases by an average 0.04 and 0.13 respectively. Adding description (D) gives an insubstantial boost (only 0.01 points of $\kappa$). Adding a narrative (N) leads to a boost of 0.06; this is modest, but perhaps the background knowledge of LLMs (especially on public data like this) is enough that the narrative adds little beyond the query terms.

Aspects (A) give a substantial improvement in $\kappa$ against TREC assessors, $+0.21$. Topicality and trustworthiness are the two aspects we used here, but of course that are not the only aspects that might matter, and we do not claim they are the best selection; at \Bing\ we use several aspects, and measure the LLM's performance on all of these with good results. It seems likely, in fact, that it is the step-by-step nature of labelling with aspects that gives rise to these improvements rather than the particulars of the aspects themselves.

Note that this presents features in isolation, when in fact any prompt could have zero, one, two, three, four, or all five of these features at once.  The effects are not additive: for example, including both a role statement and multiple judgements improves $\kappa$ despite those features being unhelpful individually. The best-performing prompt in Table~\ref{tab:results-trecrobust} is of the form ``-DNA-'', which is expected from this feature-level analysis.

\subsection{Effect of paraphrasing prompts}
\label{sec:paraphrase}

We have seen that LLM peformance varies considerably as the prompt is varied, even when the task and the input data are fixed. This raises a question: how sensitive is the LLM not just to coarse prompt features, such as asking for aspects, but to quirks of phrasing? In other words, if we rephrased ``assume that you are writing a report'' to ``pretend you are collecting information for a report'', or to ``you are collecting reading material before writing a report'', would the labels change? If so, then our LLM is highly sensitive to such apparently trivial considerations. That would mean that, first, the results above are only representative of a wide range of possible performance; and second, any serious attempt to use LLMs at scale needs to explore a large and unstructured prompt space.

To test this, we took the ``-DNA-'' prompt---the best above---and generated 42~paraphrases by rewriting the text ``Given a query and a web page \dots\ Otherwise, mark it 0'' and by rewriting the text ``Split this problem into steps: \dots\ Produce a JSON array of scores without providing any reasoning''. Figure~\ref{fig:paraphrases} gives some examples.

\begin{figure}
  {
\small\sffamily

\begin{tabular}{p{0.95\columnwidth}}
\textbf{Original, $\mathbf{\kappa=0.64}$} \\

Given a query and a web page, you must provide a score on an integer scale of 0 to 2 with the following meanings:

2 = highly relevant, very helpful for this query

1 = relevant, may be partly helpful but might contain other irrelevant content

0 = not relevant, should never be shown for this query

Assume that you are writing a report on the subject of the topic.  If you would use any of the information contained in the web page in such a report, mark it 1.  If the web page is primarily about the topic, or contains vital information about the topic, mark it 2.  Otherwise, mark it 0. \dots

\\ 

\textbf{Paraphrase, $\mathbf{\kappa=0.72}$} \\

Rate each web page for how well it matches the query, using these numbers: 0 = no match, 1 = some match, 2 = great match. Think of writing a report on the query topic. A web page gets 2 if it is mainly about the topic or has important information for the report. A web page gets 1 if it has some information for the report, but also other stuff. A web page gets 0 if it has nothing to do with the topic or the report. \dots
\end{tabular}
}
  \caption{Examples of paraphrased prompts (extracts), based on prompt format ``-DNA-'' (description, narrative, and aspects).}
  \label{fig:paraphrases}
\end{figure}

\begin{figure}
    \centering
    \includegraphics[width=\columnwidth]{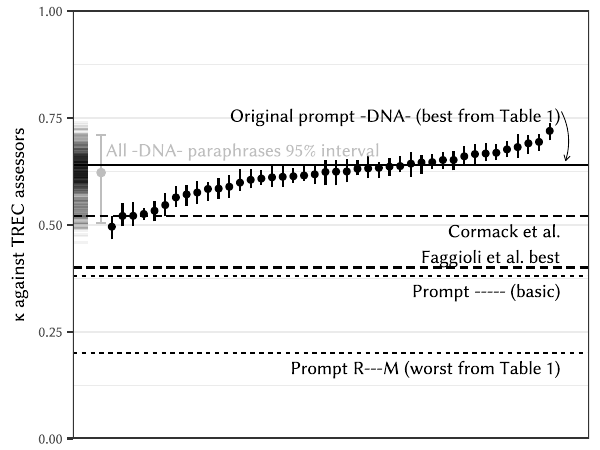}
    \caption{Variation in Cohen's $\kappa$ between LLM labels and human labels, over a stratified sample of 3000 documents from TREC-Robust, as we paraphrase the prompt.
    }
    \label{fig:kappa-variation}
\end{figure}

Figure~\ref{fig:kappa-variation} shows the resulting spread of label quality, measured again as Cohen's~$\kappa$ against the labels from TREC assessors and across our stratified sample of \num{3000} documents.  Each paraphrase is represented by one dark line, showing the mean $\kappa$ and a 95\% confidence interval derived from 20~bootstraps over documents. There is a large range, from $\textrm{mean }\kappa = 0.50$ (moderate agreement) to $\textrm{mean }\kappa = 0.72$ (substantial agreement, and better than the reference values cited above \cite{agarwal19good,castillo06spam,cormack98efficient,faggioli23perspectives,hersh94ohsumed}). The empirical 95\% confidence interval, over all bootstraps and all paraphrases, is 0.50--0.71 (plotted at the left-hand edge of Figure~\ref{fig:kappa-variation}). In contrast to \citet{wang23camels}, we saw no consistent or practical effect due to prompt or document length.

This is a wide range from a single prompt design, and from Figure~\ref{fig:paraphrases} it is not at all apparent which versions would score higher or why. The outsized effect of simple paraphrases has been observed in other domains as well \cite{zhang22tempera,zhou22large}.
This leads to two observations. First, the measured performance of any prompt---including those in Table~\ref{tab:results-trecrobust}---should be taken as a single sample from a wider range of potential performance. Small tweaks to the wording could result in noticeably different performance, even without any changes to the prompts' overall design. Second, it is prudent to fix an overall design, and then explore rephrasing and other options.
We note work by \citet{pryzant23automatic}, \citet{yang23optimizers}, \citet{zhou22large}, and others that suggests alternatives for fine-tuning prompts.

\subsection{Effect of document selection}

Given the different performance of the different prompts, and indeed the different paraphrases, it is tempting to choose the best-performing variant and commit to using it for future labelling. This of course carries a risk: performance on these topics and documents might not predict performance on other, unseen, topics and documents. The conventional guard against this is a train:test split. Here, we can interpret ``training'' as the choice of prompt, and we used repeated splits to understand the risk of choosing the best variant. For each of 1000~iterations, we randomly split our 3000 TREC and LLM labels into two sets of 1500 documents. We measured $\kappa$ for each prompt (or paraphrase) over the first 1500, noted the best performer (highest $\kappa$), and measured again on the second~1500.

The results were consistent. When scoring prompts (Table~\ref{tab:results-trecrobust}), in all 1000~iterations the best-performing prompt on the first split also beat the baseline ``-{}-{}-{}-{}-'' on the second split. That means that, starting from the baseline prompt, if we chose an alternative because it was the best improvement on one set of documents, we can be almost certain that prompt would still be an improvement on another set. In 829/1000 first splits, the best-performing variant was -DNA-, which is again consistent with the above but also suggests the choice is reliable. (The next best performer was -{}-NA-, 139 times out of 1000; in practice these two prompts are very similar.)

Looking at the 42~paraphrases of Figure~\ref{fig:kappa-variation}, in 989/1000 iterations the best-performing paraphrase on the first 1500 documents still beat the initial -DNA- prompt on the second~1500. The best-performing paraphrase was again consistent: variant~\#13 had the highest $\kappa$ on the first split in 838/1000 iterations. This is marginally less consistent than the choice of overall prompt design.

These observations suggest that while performance is variable, there is little chance of regret. That is, if we start with a baseline prompt and generate variants (e.g. by adding features or by paraphrasing) and choose to switch to the best variant, that is a safe choice. If we choose the best variant on some set of documents, performance on unseen documents will almost never turn out to be worse than the baseline.

\subsection{Query difficulty and run effectiveness}

Document labels themselves are not the goal of most evaluations. Instead, we typically map these labels to numeric values
and then use a metric such as average precision to aggregate to scores for each query and run. The scores for queries let us investigate instances where we do badly, meaning where there is scope for improvement; the scores for runs let us choose which combination of algorithms and parameters performs the best overall.

Accordingly, another way to judge a labelling scheme is by whether (under some metric) it gives the same ranking of queries or runs. If we swapped labelling schemes, would we still identify the same queries as hard? Would we still identify the same runs as top performers?

In Table~\ref{tab:queries-runs} we report the consistency of query and run rankings as we switch from human-assigned to LLM-assigned labels. In each case we score all the queries with one metric---e.g.\ P@10---based on TREC's human labels, and score them again based on our LLM labels. (We collected additional labels so that every document retrieved to depth 100, in every run, was labelled with prompt -DNA- \emph{except} those which were never labelled at TREC. For consistency with TREC, we assume these unlabelled documents are not relevant.)
This gives two rankings of queries. The consistency between these rankings is measured with RBO, normalised so that a score of~0 represents an inverted order and a score of~1 represents an identical ordering. We assume an experimenter would be willing to look at the worst ten queries, so set $\phi=0.9$.

The exercise is repeated for all 110~runs, assuming we want to find the best three or four runs ($\phi=0.7$). Since runs from the same group are likely very similar, we also repeat the exercise for the best run for each group---this simulates choosing the best approach (or perhaps vendor), rather than the best parameter settings. Again we assume we want to find the best three or four.

\begin{table}
  \centering
  \caption{Consistency of rankings on LLM labels compared to human labels, replicating all qrels in TREC-Robust to a depth of~100.
  }
  \newcommand{\ctwo}[1]{\multicolumn{2}{c}{#1}}
\begin{tabular}{rS[table-format=1.2]S[table-format=1.2]S[table-format=1.2]S[table-format=1.2]S[table-format=1.2]S[table-format=1.2]}
                    & \ctwo{Hardest queries}    & \ctwo{Best runs}                      & \ctwo{Best groups} \\
                    & {RBO} & {$\tau$}          & {RBO} & {$\tau$}          & {RBO} & {$\tau$}   \\
\cmidrule(lr){2-3}\cmidrule(lr){4-5}\cmidrule(lr){6-7}
P@10                & .40   &  .43              & .79   & .82               & .97   & .86 \\
RBP@100, $\phi=0.6$ & .42   &  .44              & .63   & .86               & .91   & .80 \\
MAP@100             & .48   &  .42              & .50   & .77               & .58   & .65 \\
\end{tabular}
  \label{tab:queries-runs}
\end{table}

The consistency of rankings, in all three cases, depends on the metric being used: ordering by MAP is more consistent for queries, and ordering by average P@10 is more consistent for runs and groups. Group-level rankings are more consistent than runs or queries, no matter the metric. It is harder to be consistent when ranking 250~queries than when ranking 110~runs or 14~groups, and small perturbations make a larger difference in ranking since many queries have similar scores. Nonetheless we see that for any problem and choice of metric, labels from LLMs lead to overall rankings which are at least similar to those from human labels, and our imagined experimenters would make similar choices. For example, under all metrics the top three runs are the same; the top five groups are consistent under P@10, the top three under RBP@100, and three of the top four under MAP@100.
The worst-performing query is the same under TREC or LLM labels for P@10 and RBP@100, and two of the top three are the same under MAP@100.

Of course perfect agreement is unlikely even with humans labelling. By way of comparison, \citet{voorhees98variations} reports $\tau=0.94$ across runs, using labels from different assessors.
This is on a different data set, with correspondingly different judgements (and only 33~runs), but give a rough upper bound for how consistent runs could ever be.  \citet{faggioli23perspectives} demonstrate $\tau$ from 0.76 to 0.86 on TREC Deep Learning data, again under slightly different circumstances (notably, shorter documents and fewer runs). We see $\tau$ from 0.77 (MAP@100) to 0.86 (P@10) for our 110~runs with full documents.
Given the $\kappa$ and AUC figures in Table~\ref{tab:results-trecrobust}, this is at least promising and plausibly as good as most human labellers.

\subsection{Observations}
 
We see somewhat better results than those reported by \citet{faggioli23perspectives}, particularly in agreement on the raw labels ($\kappa$). There are at least two factors at work. First, we are using a more capable model (GPT-4 with local modifications, compared to stock GPT-3.5); and second, our prompts are based on our experiences in Bing, and relatively long, whereas those of \citeauthor{faggioli23perspectives} are simpler. Even small wording changes can make a difference (Figure~\ref{fig:kappa-variation}), and selecting prompt features makes a bigger difference still (Table~\ref{tab:results-trecrobust}). Again, this demonstrates that time spent on this configuration---which is comparable to time spent on instruments and instructions for crowd or in-house workers---can pay dividends.

These results show that LLMs are competent at labelling---at the minimum, with GPT-4 and in the TREC-Robust setting. The labels are as close to those from humans as we could expect, given the disagreement between people to begin with, and we can reasonably consistently identify the hardest queries, best runs, and best groups.

We now turn to LLM labelling at scale, in the context of a running search engine, where LLMs have proved not just more efficient but more accurate than the status quo.
\section{Web search at \Bing}
\label{sec:web}

The results above are on one TREC corpus,
with labels from trained assessors working over simulated information needs. At \Bing\ we have also seen good results with our web corpus, queries from real use, and labels from searchers with real needs.

\subsection{Experience with LLMs}

At \Bing\ we have made heavy use of crowd workers, for many years, to scale to the number of labels, languages, and markets we need. Despite systems for detecting low quality labels and workers, this scale has come at a cost of  biases, mistakes, and adversarial workers.

\begin{table}
\caption{Labelling schemes compared. ``Crowd'' are crowd workers via our in-house platform, ``LLM'' is the best-performing prompt from private experiments.
This gives an overall comparison, but depends on our particular resources, contracts, training, and other details.}
\small
\begin{tabular}{rcccc}
              & Relative &         & Relative   & Relative \\
              & accuracy & Latency & throughput & cost \\
              \cmidrule{2-5}
Employees     & +24\% & hours--days   & $\times\,^1/_{100}$ & $\times\,8$ \\
Best crowd    & +19\% & hours--days   & $\times\,^1/_{15}$  & $\times\,5$ \\
Typical crowd & ---   & hours         & $\times\,1$         & $\times\,1$ \\
LLM (GPT-4)  & +28\% & minutes--hours & $\times\,10$        & $\times\,^1/_{20}$ \\
\end{tabular}

\label{tab:overall}
\end{table}

In Table~\ref{tab:overall} we summarise our experiences, considering full-time employees (mainly scientists and engineers working on metrics); our best crowd workers, recruited and trained for metrics problems and with close oversight; our general pool of crowd workers, subject to quality control but minimal training; and our LLM models. 

In our experience LLMs do remarkably well. They have proved more accurate than any third-party labeller, including staff; they are much faster end-to-end than any human judge, including crowd workers; they scale to much better throughput; and of course are many times cheaper. This has let us measure many more results than previously, with associated gains in sensitivity (we can see smaller effects if we label more things). The end-to-end speed, also much improved, is helping \Bing\ engineers try more things and get more done.  We have been using LLMs, in conjunction with expert human labellers, for most of our offline metrics since late~2022.

\subsection{Evaluating labellers and prompts}

In \Bing's case we have found breadth preferable to depth: that is, we prefer small data for many queries to the TREC-Robust approach of more data for fewer queries. All else being equal, we also prefer queries which resemble a real web search workload rather than the invented needs of TREC-Robust.

Our gold labels are, therefore, largely gathered in situ: from employees and contractors in the context of their normal search activity, and also from feedback from the general public.  This data is collected at or close to the time of need, by people who had the need, and in view of a full SERP (including e.g. images, maps, and advertisements). These properties mean the data is very reliable: if a label says some document is good (or bad) for the search, it is almost certainly so.

Our ground truth corpus comprises queries, descriptions of need, metadata like location and date, and at least two example results per query. Results are tagged---again, by the real searcher---as being good, neutral, or bad and these tags may be reviewed by Microsoft staff prior to inclusion in our corpus.   Similar to the TREC experiments above, from this we can derive pairs of preferred and non-preferred results and then treat labelling and scoring as a binary classification problem: the preferred result should score higher than the non-preferred, for all queries and pairs of results. Again, we can use pairwise agreement to evaluate the labels. At the time of these experiments our ground corpus comprised about 2.5~million such pairs, in about ten languages and from about fifty countries.

Using three labels does conflate small distinctions (``it's a little bit better'', e.g. good vs neutral results) and large distinctions (``it's a lot better'', good vs bad results), but our ground truth corpus has distinct advantages in that we can collect preferences from real searchers in their own context, and providing a preference is easier than providing absolute labels \citep{carterette08here}.

Our ground truth corpus gives us an evaluation which is independent of the labels from third-party judges.
In particular, by measuring against searcher-generated labels we can identify cases where the model is more accurate than third-party human judges; if we only had third-party labels, we could identify labelling disagreements but not resolve them one way or the other.
For AUC scores to be useful, of course the data must represent some population of interest: at \Bing\ we stratify the triples by important result attributes (for example language, recency, authority, or topicality). This is not a uniform sample but instead lets us identify areas of particular concern.

\subsection{Monitoring the LLM system}

The results above give us a good deal of confidence that an LLM, appropriately prompted, can produce high-quality labels for at least some important aspects. As an additional safety check, every week we take a stratified sample of recent labels from the model. Those are re-labelled by trained assessors, and we monitor for shifts in disagreement rate or in patterns of disagreement; any changes are investigated by a metrics team with expertise in both the crowd and LLM processes. In practice, large changes are rare, and resolved in favour of the LLM as often as in favour of the humans.

In addition to the human oversight of our LLM based labels we have a large set of queries that we consistently relabel.  Day to day, we expect no change in this set.
This is designed to monitor the health of labelling systems and allows a rapid response to any change.

Our system therefore sits somewhere between \citeauthor{clarke23hmc}'s ``manual verification'' and ``fully automated'' options~\cite{clarke23hmc}, with the scale of automation but some control and quality assurance from manual verification.  Disagreements, and analyses of these, inform future developments of the metrics and the gold set as well as the LLM labeller.

We note, too, that although LLM labels are important to our evaluation they are only one part of a web-scale search system. Amongst other things, web search needs to account for spam, misinformation, piracy, and other undesirable material; needs to treat some topics carefully and with editorial input (health, finance, and others); and needs to account for diversity in the final ranking. Our LLM prompts do not replace any safety systems.
\section{Potential limitations and pitfalls}
\label{sec:limitations}

We should acknowledge potential limitations and negative externalities of this approach.
Language models are known to reproduce and amplify harmful stereotypes and biases \citep{blodgett2020language,bender2021dangers,bolukbasi2016man,caliskan2017semantics,gonen2019lipstick}, and we do not know the extent of these biases in relevance labelling. This 
may intensify existing representational and allocative harms from search systems~\citep{sweeney2013discrimination, noble2018algorithms}.
Other forms of bias may also manifest, 
such as under-estimating the relevance of longer documents~\citep{hofstatter2020local}.
It may be tempting to employ a variety of different prompts and underlying LLMs to address this issue,
but that may or may not have the desired effect if 
these variations exhibit similar biases.
LLM-generated labels may also vary languages, locations, and demographic groups due to disparate training data.
This may create undesirable incentives for more pervasive data collection.

Optimising towards LLM-based labels also risks
over-fitting to the idiosyncrasies of the LLM rather than improving relevance 
\citep{chrystal01goodhart,goodhart75monetary,hoskin96awful,thomas22reliance}.
Our data 
suggests this is not yet a problem---we are closer to the ground truth with LLMs than with third-party assessors---but this may change as large models play a bigger role in ranking or as web authors start optimising for LLM labels.
LLM-generated 
labels may also show bias towards rankers 
that themselves incorporate LLMs.
Alternatively, we may view the use of LLM-based labels to evaluate and train cheaper models as a form of knowledge distillation~\citep{hinton2015distilling}, where over-fitting to the teacher may be less problematic.
Interestingly, in this context the LLM-based labeller represents a new class of machine learned relevance estimators that can be augmented with assessment guidelines as side-information.

Biases may arise not just from LLMs learning spurious correlations, but due to
differences in LLM and human attention~\citep{bolotova2020people, kazai2022less}.
Whether website designers can take advantage of such biases to unfairly gain more exposure, or whether optimising towards what LLMs deem important leads to undesirable shifts and homogenisation of online content\footnote{\url{https://www.theverge.com/2019/5/28/18642978/music-streaming-spotify-song-length-distribution-production-switched-on-pop-vergecast-interview}}, are also important questions.

Lastly, the ecological costs of these LLMs are still heavily debated~\citep{bender2021dangers, patterson2021carbon, bommasani2021opportunities, wu2022sustainable, dodge2022measuring, patterson2022carbon} and an important area for further study.
\section{Concluding remarks}
\label{sec:conclusion}

Evaluating information retrieval typically relies on relevance labels, and we have several options for collecting these. Figure~\ref{fig:pareto-space} illustrates the options discussed in this paper. As experimenters, our goal is to move up and left, to greater accuracy and lower cost. Traditionally the goal has been to improve crowd labels---moving the bottom-left point higher up---and this has involved
\begin{enumerate*}[label=(\roman*)]
    \item collecting insight from real searchers,
    \item turning this into guidelines,
    \item using trusted workers to read these guidelines and generate ``silver'' labels, and
    \item giving the same guidelines to crowd workers.
\end{enumerate*}
The crowd workers are monitored against the silver labels, and improvements largely come from improving the guidelines.

Our approach is different: we collect high-quality gold labels from searchers themselves
and use these labels to evaluate and select prompts for an LLM. The labels we get from our model are high quality, and in practice are more useful than those from even trained assessors. They are of course cheaper to acquire, and easier to collect for new languages or other new context; but they are also more accurate than third-party labels at predicting the preference of real searchers. This has had a tangible effect: retraining parts of our ranker using labels from this model, while keeping all else constant, resulted in about six months' relevance improvement in a single step.

Of the options described by \citet{faggioli23perspectives}, our labelling is closest to ``human verification'',
although we do not deliberately vary the LLM's characteristics. We do retain human oversight and audit examples of LLM output, although not every label. Quality control, and indeed measuring LLM quality in general, is (as anticipated by \citeauthor{faggioli23perspectives}) difficult as in most cases our LLM is ``beyond human'' quality and we can no longer rely on third-party assessors. Our gold collection, with queries and labels from real searches and real searchers, helps a great deal but of course searchers can still be swayed by distracting captions or unreliable results. (We review every query and URL in the corpus, but this only adds another human to the loop.) Contra \citeauthor{clarke23hmc}, we do not see machine-made assessments degrading quality at all; nor do we consider them ``very expensive'', at least compared to trained annotators.

In some ways, this is an easy case: the language model was trained on web text and we are labelling web text. The notion of judging web pages is likely already encoded, although we do not have clear evidence for this. Further, the topics can be addressed in the corpus: they do not need any personal, corporate, or otherwise restricted data, nor any particular domain-specific knowledge not already found in the text. Using LLMs for labelling suggests new and more difficult applications, for example labelling private corpora where we cannot give human assessors access. From the experiments above, we cannot verify this will be effective, and this remains for future work. We have also measured our labels in part with test sets---both TREC, and \Bing's corpus---which have clear task descriptions. If we were to sample a query load from a running system, we would not have these descriptions and our labels would be less accurate. We also have a capable model: \citet{liang22helm} saw large differences from model to model over a range of tasks, although given our observations in Section~\ref{sec:results} this could also be due to model:prompt interactions. As new models emerge, they will of course need to be tested.

The results above use one particular model. As models improve, it becomes harder to measure our labels as the measures start to saturate \cite{faggioli23perspectives}. We have found it necessary to build harder gold sets over time, encoding finer distinctions to better distinguish labellers and prompts. There is no equivalent mechanism in open data sets, and this may become pressing should LLM-based labelling become common.

\vspace{\baselineskip}

\noindent It is certainly possible to use LLMs to label documents for relevance and therefore to evaluate search systems; it is possible to get performance comparable to TREC judges and notably better than crowd judges. There are many choices that make a difference, meaning we need metrics-for-metrics to distinguish a good from a bad system, as well as ongoing audits and human verification. In our experience, having true ``gold'' judgements 
makes it possible to experiment with prompt and metric design. We have found the approach productive at \Bing, and have used it for greater speed, reduced cost, and substantial improvements in our running system.

\begin{acks}
We thank David Soukal and Stifler Sun for their effort developing and testing many iterations of \Bing's LLM labelling system. Ian Soboroff kindly provided TREC-Robust judging guidelines. Dave Hedengren, Andy Oakley, and colleagues at \Bing\ provided useful comments on the manuscript.
\end{acks}

\balance
\bibliographystyle{ACM-Reference-Format}
\bibliography{abbrevs-long,llm}

\ifappendix
\appendix
\section{Further experimental results}

\subsection{LLM-vs-human confusion matrix}

Additionally to the evaluations in Section~\ref{sec:results}, we can directly compare our model-generated scores to assessor scores for each query:document pair in our stratified TREC sample. Table~\ref{tab:confusion-trecrobust} gives a confusion matrix for one prompt and all \num{3000} query:document pairs. (There are 32~such matrices, one for each set of prompt features or equivalently one for each row of Table~\ref{tab:results-trecrobust}.) We can see that in this case, the LLM is more likely to say ``not relevant'' than were TREC assessors (44\% vs 33\%), and is correspondingly inaccurate (68\% agreement with TREC, when the LLM says ``not relevant''). An LLM assessment of ``relevant'' or ``highly relevant'' however, is reliable (94\% agreement).

\begin{table}
\begin{tabular}{rccc}
&  & \multicolumn{2}{c}{Model} \\
&  & 0      & 1 or 2 \\
\cmidrule{3-4}
\multirow{2}{*}{TREC assessor}
   & \multicolumn{1}{r|}{0}      & 866 &   95 \\
   & \multicolumn{1}{r|}{1 or 2} & 405 & 1585 \\
\end{tabular}
\vspace{2em}
\caption{Results from the best-performing prompt of Figure~\ref{fig:prompt-trecrobust}---i.e. with descriptions, narrative, and aspects, prompt ``-DNA-''---over a stratified sample of the TREC Robust data. Overall, the LLM is more likely to say ``not relevant'' than were TREC assessors; an LLM assessment of ``relevant'' or ``highly relevant'' is reliable. Some qrels are missing due to unparsable LLM output, a rate of 1.6\%.}
\label{tab:confusion-trecrobust}
\end{table}

\subsection{Effect of prompt length}

Using an LLM to compare texts, \citet{wang23camels} saw an effect of prompt length---the longer the text, the more positive the LLM's assessment. We checked for similar effects in our data by modelling the LLM's \emph{signed error} as a response to prompt length. This controls for any effect of length on true relevance; if longer documents are just more (or less) likely to be relevant, then the LLM should not be penalised for reflecting this. Replicating \citeauthor{wang23camels}'s effect would require a positive effect: that is, errors should get more positive (the LLM should overestimate more, or be more optimistic) as prompts got longer.

Controlling for prompt features, we saw no substantial correlation between prompt length and signed error. Effects varied according to prompt features, with modelled score shifting between $-9\times10^{-6}$ and $1\times10^{-5}$ per character of prompt. This corresponds to only a shift in score of -0.05~to~0.06 at the median prompt length, which (although statistically significant) is of no practical significance given the MAEs of Table~\ref{tab:results-trecrobust}.

\fi

\end{document}